\documentclass{ieeeaccess}
\usepackage{cite}
\usepackage{amsmath,amssymb,amsfonts}
\usepackage{algorithmic}
\usepackage{graphicx}
\usepackage{textcomp}
\def\BibTeX{{\rm B\kern-.05em{\sc i\kern-.025em b}\kern-.08em
    T\kern-.1667em\lower.7ex\hbox{E}\kern-.125emX}}

\usepackage{hyperref}
\usepackage{cleveref}

\usepackage{color,soul}
\soulregister\cite7
\soulregister\Cref7
\setulcolor{red}
\sethlcolor{yellow}
\renewcommand\hl[1]{#1}

\usepackage{caption,setspace}
\usepackage[export]{adjustbox}

\begin{document}
\bstctlcite{IEEEexample:BSTcontrol}
\history{TODO: Date of publication xxxx 00, 0000, date of current version xxxx 00, 0000.}
\doi{TODO: 10.1109/ACCESS.2020.DOI}

\title{An In-Depth Security Assessment of Maritime Container Terminal Software Systems}

\author{\uppercase{
    Joseph~O.~Eichenhofer\authorrefmark{1},
    Elisa~Heymann\authorrefmark{1}\authorrefmark{2},
    Barton~P.~Miller\authorrefmark{1},
    and~Arnold~Kang\authorrefmark{3}
}}

\address[1]{
    Computer Sciences Department,
    University of Wisconsin - Madison,
    Madison, WI 53706 USA
    (email: eichenhofer, elisa, bart @cs.wisc.edu)
}

\address[2]{
    Departamento de Arquitectura de Computadores y Sistemas Operativos,
    Universitat Aut{\`o}noma de Barcelona,
    Bellaterra 08193 (Barcelona) Spain
    (email: elisa.heymann@uab.es)
}
\address[3]{
    Total Soft Bank Co.,
    Busan, Korea
    (email: arnold@tsb.co.kr)
}

\tfootnote{
This work is supported in part by National Science Foundation Cyber Infrastructure grant ACI-1547272, the Department of Homeland Security under AFRL Contract FA8750-12-2-0289, the University of Wisconsin-Madison, and the Agencia Estatal de Investigaci{\'o}n (AEI), Spain, the Fondo Europeo de Desarrollo Regional (FEDER) UE, under contract TIN2017-84875-P, and partially funded by EUG.
}

\markboth
{Eichenhofer \headeretal: In-Depth Security Assessment of Maritime Container Terminal Software Systems}
{Eichenhofer \headeretal: In-Depth Security Assessment of Maritime Container Terminal Software Systems}

\corresp{Corresponding author: Elisa Heymann (email: elisa.heymann@uab.es)}

\begin{abstract}
Attacks on software systems occur world-wide on a daily basis targeting individuals, corporations, and governments alike. The systems that facilitate maritime shipping are at risk of serious disruptions, and these disruptions can stem from vulnerabilities in the software and processes used in these systems. These vulnerabilities leave such systems open to cyber-attack.

Assessments of the security of maritime shipping systems have focused on identifying risks but have not taken the critical (and expensive) next step of actually identifying vulnerabilities present in these systems. While such risk assessments are important, they have not provided the detailed identification of security issues in the systems that control these ports and their terminals.

In response, we formed a key collaboration between an experienced academic cybersecurity team and a well-known commercial software provider that manages maritime shipping. We performed an analysis of the information flow involved in the maritime shipping process, and then executed an in-depth vulnerability assessment of the software that manages freight systems. In this paper, we show the flow of information involved in the freight shipping process and explain how we performed the in-depth assessment, summarizing our findings. Like every large software system, maritime shipping systems have vulnerabilities.

\end{abstract}


\titlepgskip=-15pt

\maketitle

\section{Introduction}
\label{sec:introduction}
The maritime sector is crucial to the world economy, and the computer technology that manages it is critical to its successful operation. Maritime ports in the EU handled 4.0 billion metric tons of seaborne goods in 2017, which marked an increase of 11.69\% when compared with 2009 \cite{eurostat2020}. In the US, in 2018 maritime ports collectively handled 70.7\% of America's international trade by weight \cite{aylward2016port}. Maritime shipping uses millions of containers and employs millions of people to move billions of tons of freight annually. The world economy is therefore critically dependent upon the maritime movement of cargo and containers. Consequently, the economy is also dependent upon the software systems that facilitate maritime operations.

Maritime freight transportation increasingly relies on Information and Communications Technology (ICT) to manage and optimize its operations and services. ICT makes the essential operations not only manageable but also cost effective. This technology is involved in many areas, from traffic control communications to container freight tracking to the actual movement of containers. As a consequence, there is an increased dependency on electronic communication and processes with little human interaction. In addition to these benefits, the freight ICT systems also introduce the risks of being extremely vulnerable to cyber-attack. \emph{It is important to note that these ICT systems are based largely on software that has been written specifically to support the operations of maritime freight systems.}

Freight ICT systems are large and complex, having many components used by different principals involved in the supply chain. Some of these components are used by the general customers, for example the Port Community System (PCS), to book and track shipments and exchange documents and information between public and stakeholders. Other components are intended to be used by port operators, for example the Terminal Operating System (TOS), to control container movement and storage in the maritime port. There is also a back-office management and integration system, which allows companies to manage, link, and share internal processes with suppliers and customers. Attackers can take advantage of the complexity of this diverse collection of software. For example, in 2013 drug traffickers recruited hackers to breach the ICT systems that controlled the movement and location of containers in the Belgian port of Antwerp, managing to reroute (for two years) containers carrying drugs, guns, and cash \cite{bateman2013police}. In 2017, Maersk was hit by the devastating NotPetya cyber attack. This attack disabled their entire IT infrastructure, affecting the company's operational capacity for months and costing Maersk around \$300 million \cite{novet2017maersk}. Other recent maritime cyber attack examples include two attacks in 2018: one on the Port of Barcelona (Spain) and another on the Port of San Diego (USA) \cite{esage2018barcelona, vangrove2018sandiego}.

The software that manages and controls freight transportation systems must be hardened against cyber-attacks. Disruption or unavailability of these ICT systems could have disastrous consequences in cost and availability of goods. Attacks against vulnerabilities in the software can lead to a wide range of consequences. These consequences include disruption of service, shipment of cargo to unintended destinations, threat to human lives (for example, by remotely controlling the twistlocks of a container spreader to release it over a person), and operation of seaport machinery by unauthorized users. Therefore, there is a critical need to ensure the robustness of the ICT systems and to secure them against cyber-attacks.

This research represents the first in-depth analysis of a software system that controls maritime shipping. \hl{As of June 2020, the software assessed is used in almost 100 container terminals worldwide, therefore our assessment contributed to make our world a bit more secure.} While there have been significant efforts at assessing risk in such transportation environments, and even external penetration tests on port facilities, the software itself is at risk.\hl{Approaches like attack and mitigation trees \cite{frydman2014risk} are most useful when they are used at design time in the software development life cycle. In our work we assess already-implemented software, so cannot assume that such secure design work was done (and, in most cases, it has not been done). In addition, attack trees are best at finding vulnerabilities for which a tree has been provided; they are not designed to find new types of vulnerabilities.} For commodity software, like the Windows or Linux operating systems,  the risk of exploitation is shared by many user communities. For maritime shipping (as in many other transportation sectors), the user community is smaller, the risk more focused, and the consequences of a breach substantial. It is essential that there be:

\begin{enumerate}
    \item A global recognition of the risk of not assessing the software in depth
    \item The willingness for software providers to allow scrutiny of their software
    \item Resources available to accomplish the in-depth software assessments
    \item Transparency and reporting for the results of such assessments
    \item Training available for the transportation software practitioners to learn the skills of building secure systems
    \item Regulations that capture the requirements for improved software security
\end{enumerate}

Our effort represents an important bridge between best practices in academia and a world leader in container terminal software. In our experience, it takes courage and a leap of faith to expose your commercial software to such detailed evaluation. However, the benefits of such an evaluation can be huge, including both a significant improvement in operational security and an increased confidence in the systems by the stakeholders depending on the software.

In the next section, we review the most closely related research in this area. In \Cref{sec:shipping}, we present an overview of the surprisingly intricate flow of information that takes a container from the exporter to the importer. In \Cref{sec:assessment}, we then describe the in-depth software vulnerability methodology, called First Principles Vulnerability Assessment (FPVA), that we used in this effort. In \Cref{sec:results}, we present the results of our FPVA assessment, including descriptions of the vulnerabilities found and the remediation strategies used.

\Figure[t]()[scale=0.4]{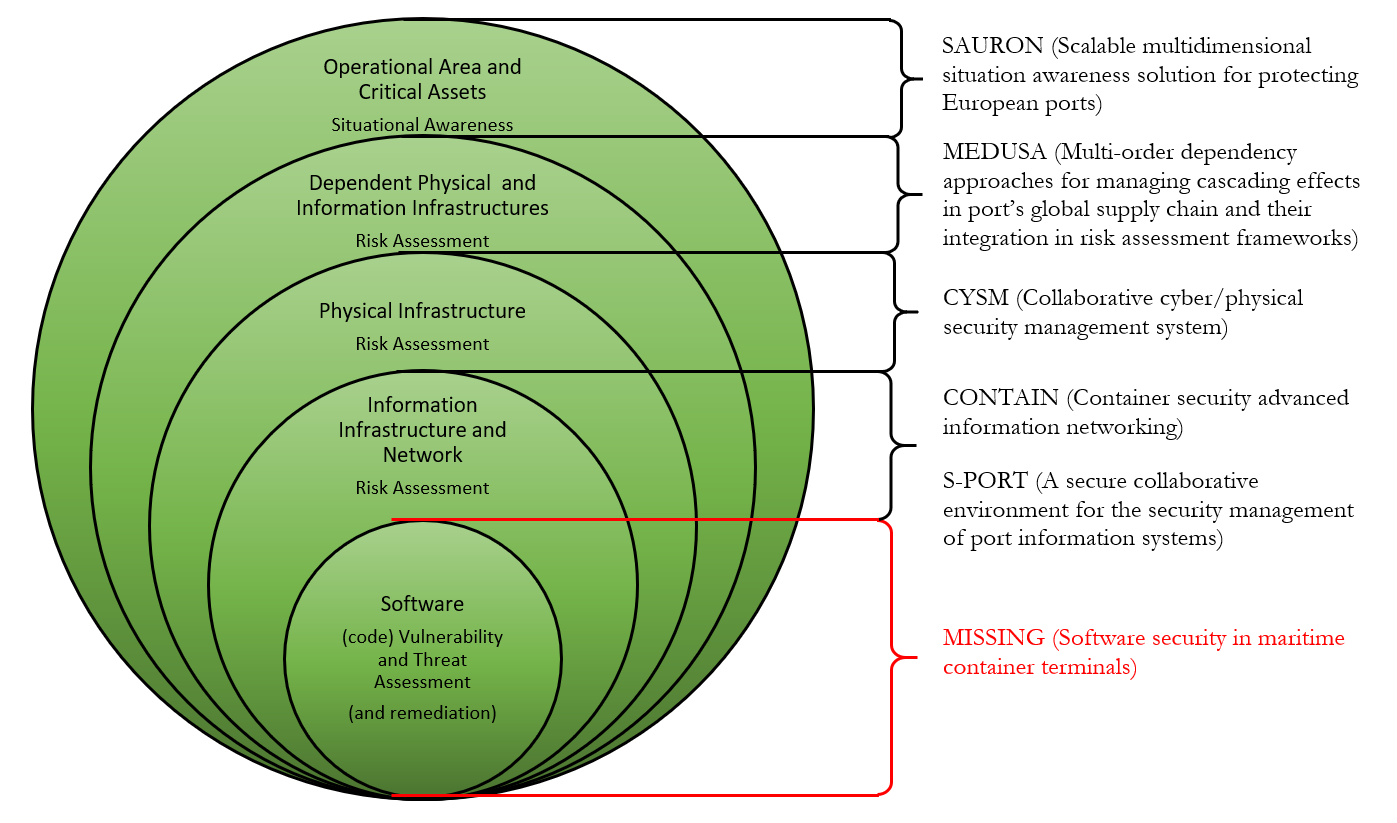}{
    Cyber-Physical Security Efforts
    \label{fig:efforts}
}

\Figure[t!]()[scale=0.9]{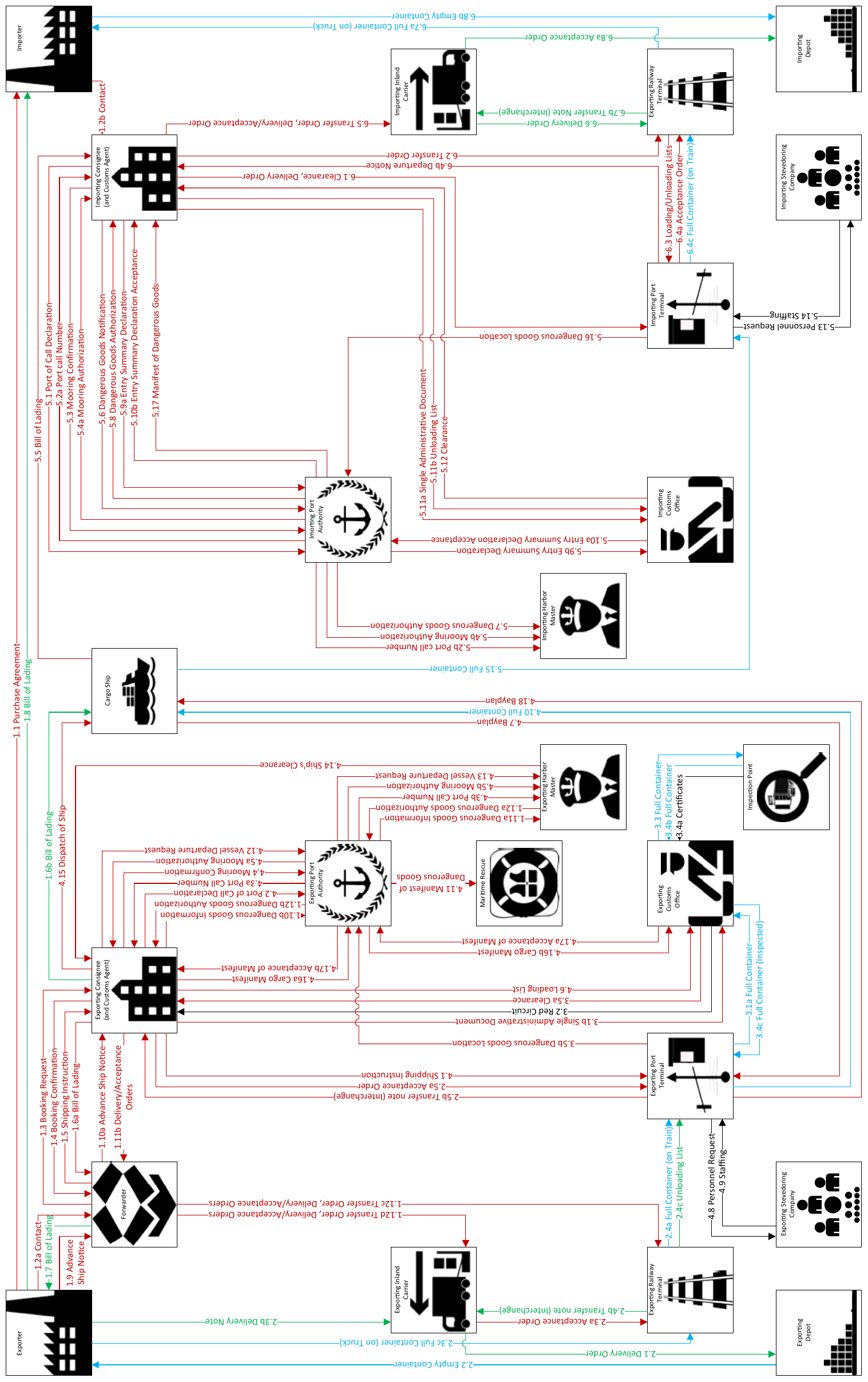}{
    Shipping Logistics Data Flow. \\
    \hl{Note that \Cref{fig:1booking} -- \Cref{fig:6delivery} show each area of this figure at a larger scale.}
    \label{fig:0full}
}

\section{Related Work}
\label{sec:related}
There has been an increasing awareness of port security in the past decade. Nevertheless, assessment of the security of maritime freight systems (in both the E.U. and U.S.) has faced two significant limitations. First, while existing studies have been directed at taking the important first step of identifying risks, they have not taken the critical and expensive next step of actually identifying the vulnerabilities present in the ICT systems. Second, these studies have focused on overall port operations. While such overviews are important and have resulted in overall recommendations for policy change, they have not provided a detailed evaluation of security issues in the ICT systems that control these ports.

In this section, we review related work in the areas of risk assessment in container seaports, focusing on its relationship to in-depth software assessment of maritime freight ICT systems.

There have been several efforts to address the risk assessment of seaports. Current efforts for risk assessment for maritime security are summarized in \Cref{fig:efforts}.

SAURON is an ongoing European project whose goal is to develop a platform for port operators to have physical, cyber, and hybrid situational awareness \cite{sauron2017}. SAURON is investigating the prevention, detection, response, and mitigation of physical and cyber threats to ports.

Previous European Projects like MEDUSA and MITIGATE focused on assessing risks in the maritime supply chain and port/maritime systems \cite{papastergiou2018design, mitigate2015}. MEDUSA concentrated on the port IT infrastructure at the supply chain level, while MITIGATE concentrated at the asset level. These approaches are intended to quantify risk, but not whether a vulnerability in the code exists, where it exists, or how it might affect the higher-level spheres (physical assets, networks, information infrastructure).

Existing security standards, best practices, maritime regulation, and risk assessment methodologies and tools fail to adequately address the specific needs of port authorities \cite{imo2002isps, imo2002ammend}. Researchers in the S-Port project developed a prototype software platform consisting of a collaborative environment to host security management services and guide commercial ports to monitor and self-manage their port ICT security \cite{polemi2013sport}. Safety standards and regulations were identified (specifically in ISO 27001 and ISPS Code), and then actions were taken to address some specific security management needs of port ICT systems. The architecture of the S-Port platform incorporates various collaborative tools, which are focused on high-level risk assessment \cite{ntouskas2010sport}. 

Historically, physical security has been the main emphasis when thinking about port security; the various seaports standardization bodies did not specifically reference ICT/Cyber-security in their memoranda \cite{polemi2012sport}. Most of the existing freight seaport security standards and methodologies concentrated only on the physical security of the ports (i.e., safety concerns) \cite{ntouskas2012collaborative}.

The International Maritime Organization (IMO) developed guidelines for maritime cyber risks as the basis for future regulation in the maritime and seaport sector. During the IMO's Maritime Safety Committee (MSC) session held in June 2017, the Committee approved MSC.428(98) Maritime Cyber Risk Management in Safety Management Systems \cite{imo2017risk}. Following MSC.1/Circ.1526, which was superseded by MSC-FAL.1/Circ.3, the resolution affirms that approved safety management systems should take cyber risk management into account, considering also confidentiality for certain aspects of cyber risk management \cite{imo2017guidelines, imo2016intirim}. The updated guidelines provided recommendations to safeguard shipping from current and emerging cyber threats and vulnerabilities. That document acknowledges that vulnerabilities can result from inadequacies in design, integration and/or maintenance of systems, as well as lapses in cyber discipline. In particular, they describe five elements to identify and manage cyber risks: (1) \textit{identify}, (2) \textit{protect}, (3) \textit{detect}, (4) \textit{respond}, and (5) \textit{recover}. \hl{These steps complement a kill chain approach to vulnerability identification \cite{higgens2013lockheed}. The kill chain is the sequence of steps taken by an attacker to accomplish an attack; if the defender breaks (defeat) any of those steps, then the attack is prevented. To use a kill chain, you first need to identify the elements in the chain (step 1, identify) and then remove one or more of those elements (step 2, protect). So, kill chains can be thought of as elaborating on and complementing the first 2 steps.}

Our in-depth software vulnerability assessment activities (described in Sections \ref{sec:assessment} and \ref{sec:results}) directly addresses the first three of these elements by:

\begin{enumerate}
    \item \textit{Identifying} the parts of the software that are of greatest risk
    \item \textit{Protecting} the software by removing the vulnerabilities
    \item \textit{Detecting} potential points of attack before they can be exploited
\end{enumerate}

The 2017 update to the guidelines further emphasized the importance of what is in the 2016 edition.

Since port ICT systems face combined physical and cyber threats, a holistic risk assessment methodology for these infrastructures should combine the analysis of physical and ICT aspects. For example, using MSRAM \cite{downs2017maritime} and CMA \cite{kang2009overview} for physical risk assessment, and using CRAMM \cite{yazar2002qualitative}, OCTAVE \cite{alberts2001octave}, or current standards such as ISO27005 and ISO27032 \cite{iso2011, iso2012} and NIST-SP 800-30 \cite{nist2012} for ICT risk assessment.

While awareness of cyber risks is steadily increasing in the maritime sector, we need to go beyond risk assessment to the actual evaluation of software systems that operate in this environment. The first step to an in-depth assessment of the software that controls maritime freight shipping consists of understanding the software involved. There cannot be a serious cybersecurity analysis without considering the software. For that purpose, we investigated the maritime shipping process and documented all the transactions (both electronic and in paper) involved. This documentation is detailed in the next section.

\section{Understanding Shipping Logistics}
\label{sec:shipping}
The process by which a shipping container carries goods from an exporter in one country to an importer in another can be viewed as a series of document and communication transactions. To begin our evaluation of these transactions, we used documentation prepared by the Port of Valencia, Spain \cite{montfort2012}.

\Cref{fig:0full} shows the communications/transactions involved in shipping logistics. Due to the large and complex nature of freight logistics, it is beneficial to approach the process in stages. For the purposes of this paper, there are six such stages: booking, forwarding, outbound customs, outbound shipping, inbound shipping, and delivery. To better visualize these stages, the transactions involved in each stage are shown in \Cref{fig:1booking} through \Cref{fig:6delivery}. Each arrow represents a transaction of paper document (green), digital document (red), container movement (blue), or unspecified communication (black). Transactions are chronologically numbered. Simultaneous transactions in the same figure share the same number and are identified by letter.

\subsection{Booking}
Several booking-related documents must be created and exchanged before the container can be moved. In this section, parenthesized numbers refer to edges in \Cref{fig:1booking}. The importer and exporter first agree on the goods to be purchased and shipped (1.1). For the sake of simplicity, we do not show the importer in this figure. The exporter contacts the freight forwarder (1.2a) who will negotiate shipment with the consignee that operates in the desired seaport (1.3, 1.4, and 1.5). A Bill of Lading is created by the consignee and given to the cargo ship, forwarder, exporter, and importer (1.6a, 1.6b, 1.7, and 1.8). When the exporter is ready to ship, it sends an advance ship notice to the forwarder who sends it to the consignee (1.9 and 1.10a). The consignee sends delivery and acceptance orders to the forwarder (1.11b) who sends them to the inland carrier and railway (1.12d and 1.12c). If the shipment is to contain any dangerous goods, the consignee reports them to the port authority (1.10b). When the port authority and harbor master approve the goods, authorization is recorded and given to the consignee (1.11a, 1.12a, and1.12b).

\Figure[t!]()[scale=1.3, fbox, angle=90, origin=c]{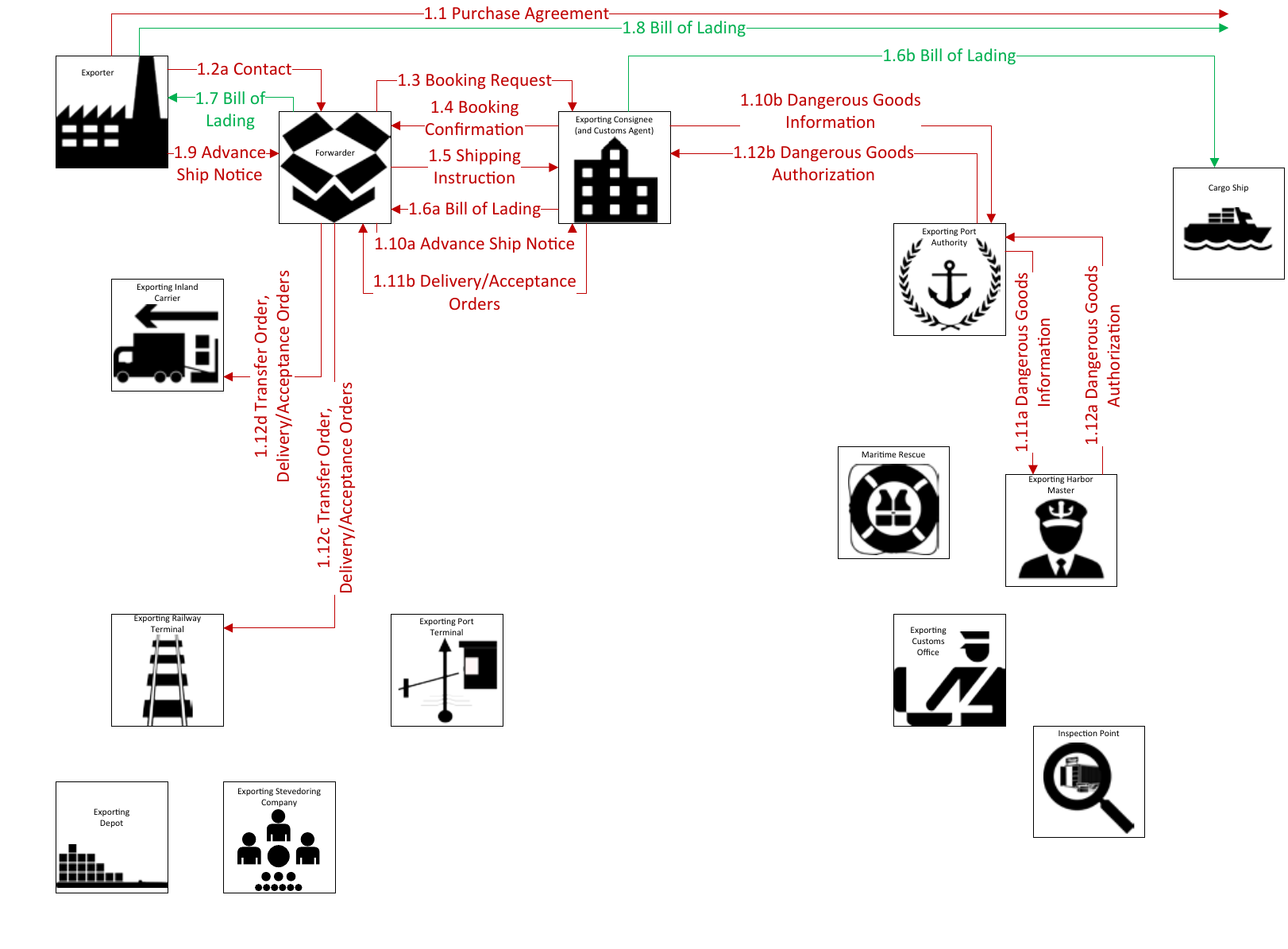}{
    Booking Logistic Data Flow
    \label{fig:1booking}
}

\subsection{Forwarding}
Once booking documents are in place, the goods will be forwarded to the seaport. Parenthesized numbers in this section refer to edges in \Cref{fig:2forwarding}. The inland carrier first takes the delivery order to the depot at the seaport to receive the consignee's container and takes the empty container to the exporter (2.1 - 2.2). The container is packed and sealed in the presence of a representative of the exporter who signs a delivery note and gives it to the carrier (2.3b). The carrier takes the full container and an acceptance order to the railway terminal (2.3a and 2.3c). The carrier is given a transfer note to document the exchange (2.4b). The railway operator loads and sends the container to the port terminal along with an unloading list that documents the goods (2.4a and 2.4c). The consignee sends to the terminal an acceptance order, and the terminal sends the consignee a transfer note to document the interchange (2.5a and 2.5b).

\Figure[t!]()[scale=1.3, fbox, angle=90, origin=c]{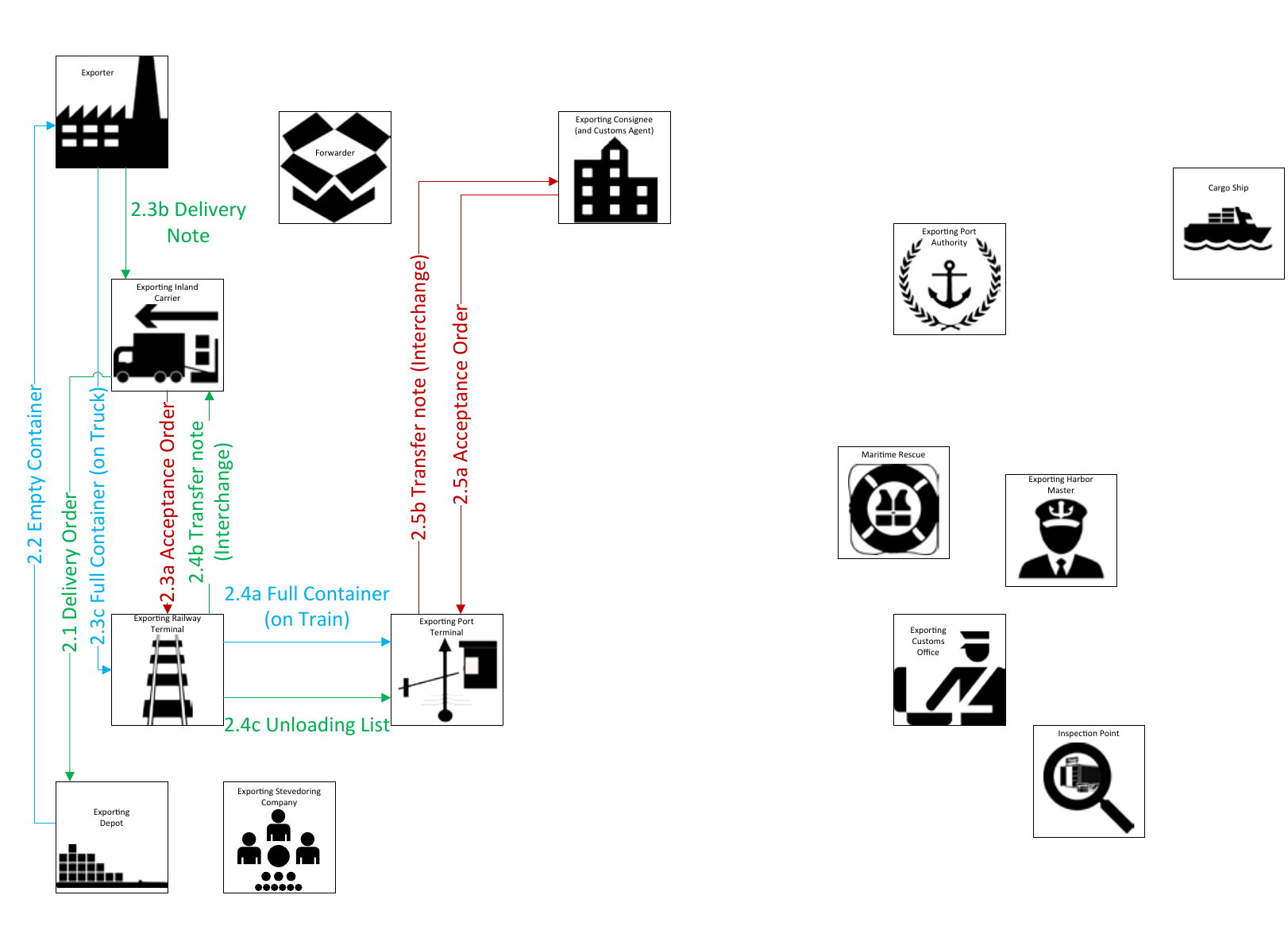}{
    Forwarding Logistic Data Flow
    \label{fig:2forwarding}
}

\subsection{Outbound Customs}
Many containers are subject to customs clearance and/or inspection once they arrive at the seaport. Edges in \Cref{fig:3customs} are referenced by parenthetical numbers in this section. The container is taken to a checkpoint run by the customs office (3.1a). Customs declarations are sent by the consignee in the form of a ``Single Administrative Document'' to the customs office at the port (3.1b). If the container is to be inspected, a ``red circuit'' is initiated (3.2). The container is moved to the inspection site (3.3), certified, and returned to the customs office and port terminal (3.4a, 3.4b, and 3.4c). Clearance documentation is sent to the consignee (3.5a). If the container contains any dangerous goods, they are reported to and tracked by the port authority (3.5b).

\Figure[t!]()[scale=1.3, fbox, angle=90, origin=c]{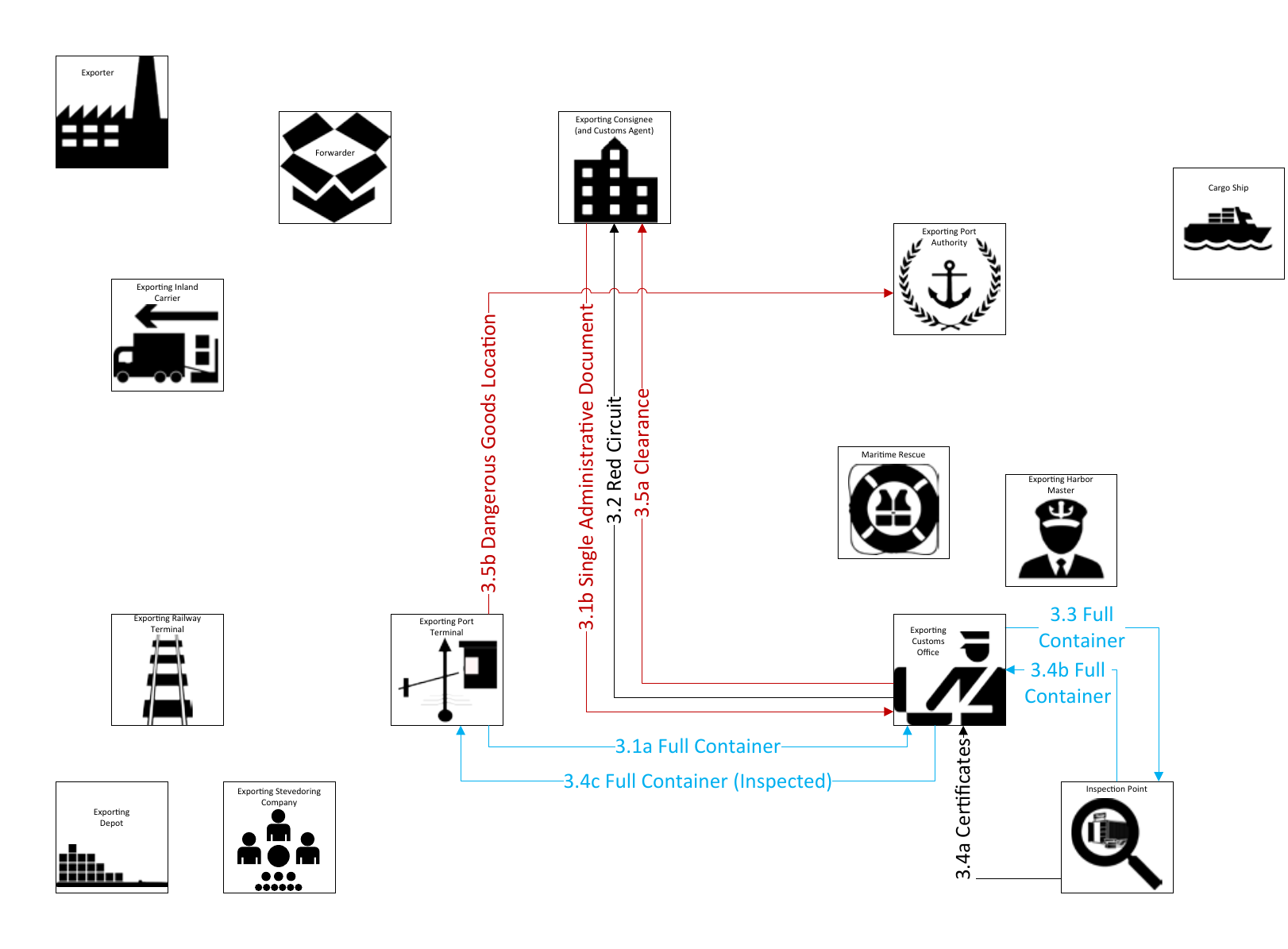}{
    Outbound Customs Logistic Data Flow
    \label{fig:3customs}
}

\Figure[t!]()[scale=1.3, fbox, angle=90, origin=c]{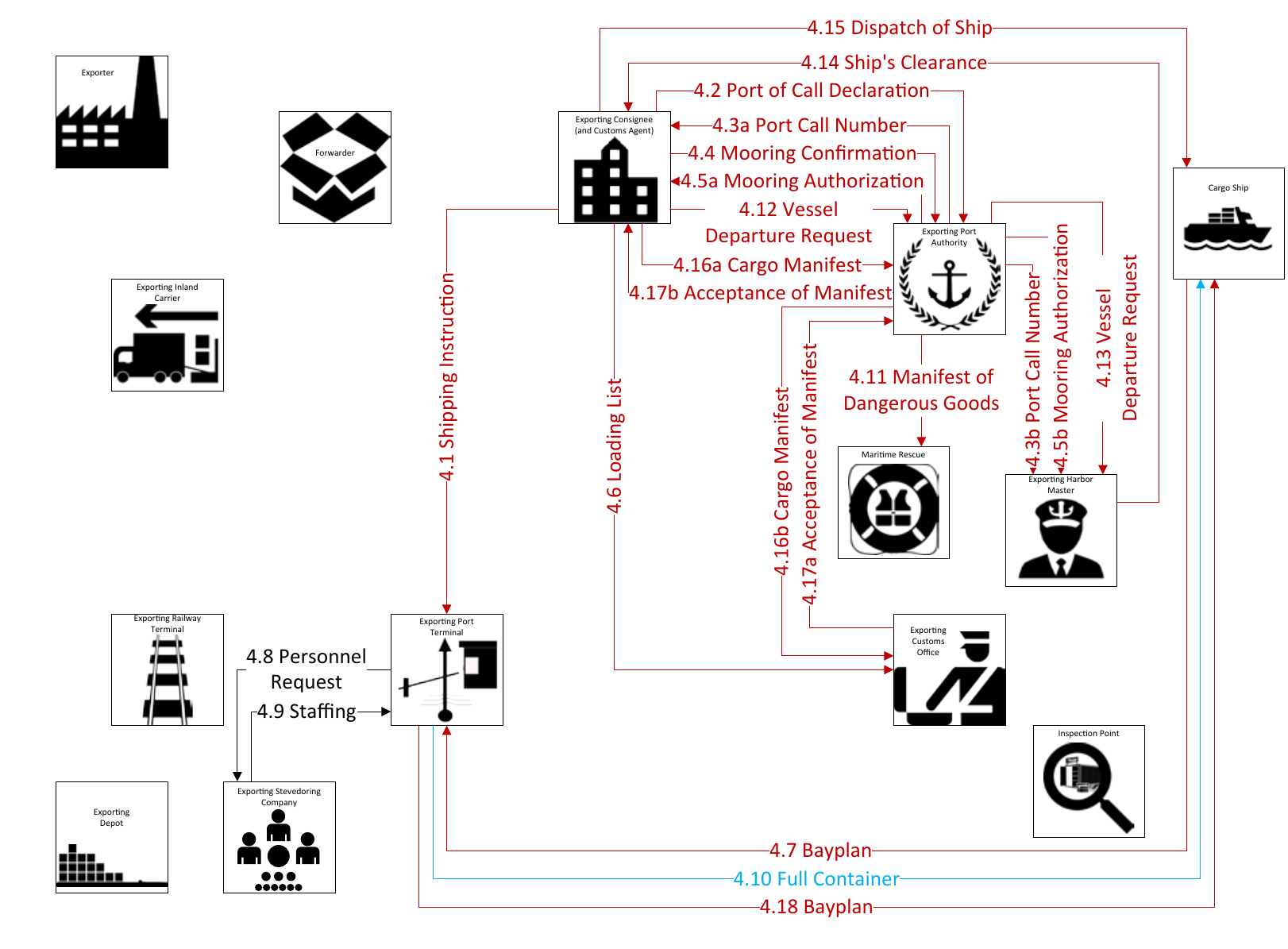}{
    Outbound Shipping Logistic Data Flow
    \label{fig:4oshipping}
}

\subsection{Outbound Shipping}
With the container certified and available at the terminal, arrangements must be made for its loading and shipment out of the port. In this section, parenthetical numbers are references to single edges in \Cref{fig:4oshipping}. After sending shipment instructions to the terminal (4.1), the consignee sends and receives authorizing documents to the port authority for the cargo ship to dock (4.2, 4.3a, 4.4, and 4.5a), some of which are sent to the harbor master for record and reference (4.3b and 4.5b). The consignee must also report to the customs office a loading list for record of the goods (4.6). The docked ship then sends its bayplan to the terminal (4.7), where arrangements are made for the ship to be unloaded and loaded by stevedores (4.8, 4.9, and 4.10). If any dangerous goods are loaded, they are reported to maritime rescue authorities for tracking (4.11). Once loading is complete, the consignee makes a request to the port authority to embark (4.12), and notifies the ship after it is authorized by the port authority and harbor master (4.13, 4.14, and 4.15). The consignee sends a cargo manifest to the port authority (4.16a), which reviews it with the customs office before documenting its acceptance (4.16b, 4.17a, and 4.17b). An updated bayplan is sent back to the ship as it departs (4.18).

\subsection{Inbound Shipping}
The process of shipment into the receiving port begins as the cargo ship nears it. Parenthetical numbers in this section refer to edges in \Cref{fig:5ishipping}. When the cargo ship approaches the receiving port, the consignee arranges for authorization from the port authority to dock (5.1, 5.2a, 5.3, and 5.4a). The port call number and mooring authorization are sent to the harbor master for record (5.2b and 5.4b). Dangerous goods must be reported to and authorized by the port authority and recorded by the harbor master (5.6, 5.7, and 5.8). The consignee sends an entry summary declaration to the port authority which forwards it to the customs office (5.9a and 5.9b). The customs office accepts the declaration (5.10a), and the consignee is notified (5.10b). A Single Administrative Document is sent to the customs office along with an unloading list (5.11a and 5.11b). Once customs clearance is granted (5.12), the port terminal arranges for stevedores to unload and load the ship (5.13, 5.14, and 5.15). Locations of dangerous goods are reported to the port authority (5.16), and a manifest of them are sent to the consignee (5.17).

\Figure[t!]()[scale=1.3, fbox, angle=90, origin=c]{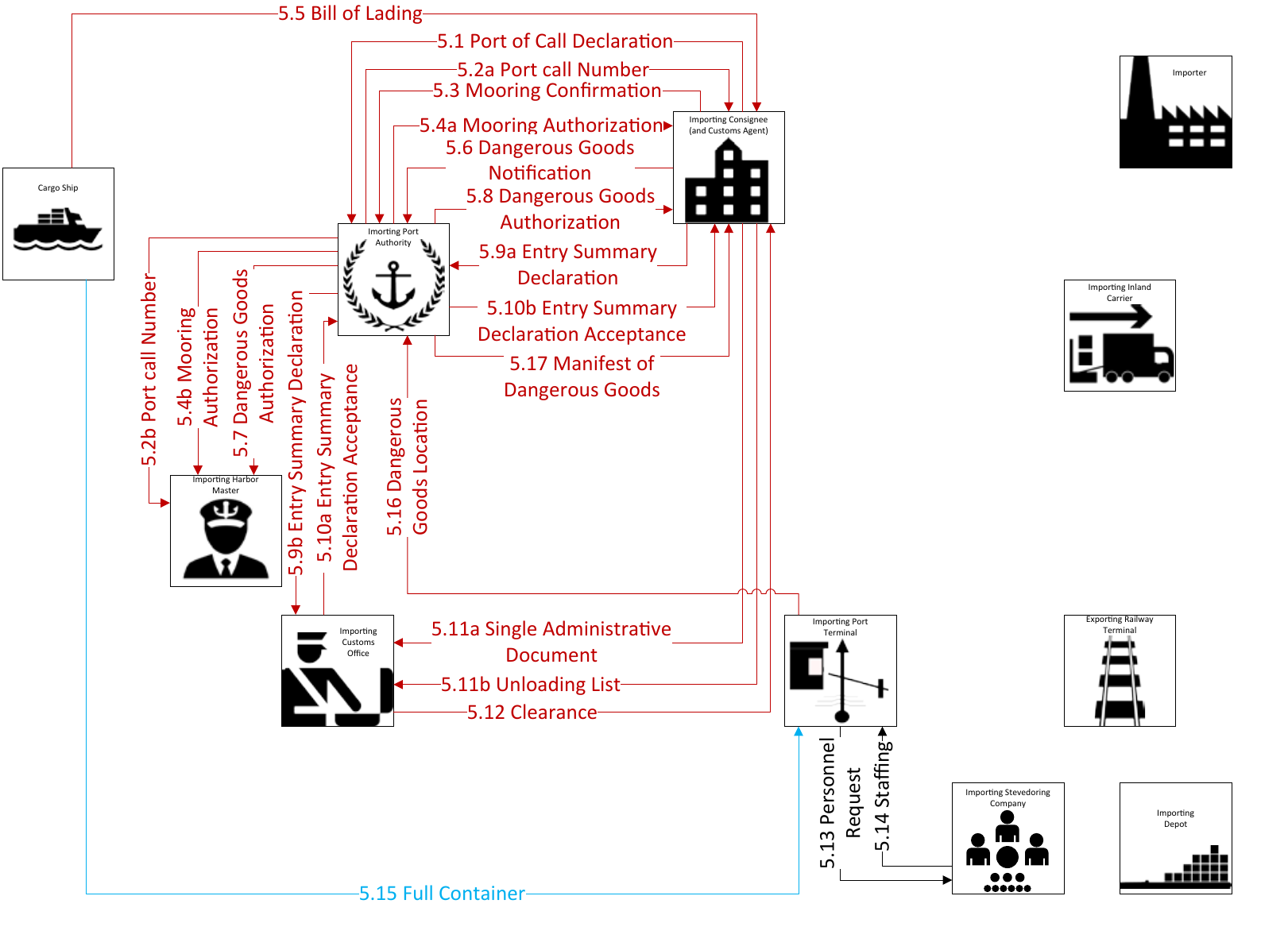}{
    Inbound Shipping Logistic Data Flow
    \label{fig:5ishipping}
}

\subsection{Delivery}
The final stage of the process is to move the full container from the port, deliver the goods to the importer, and return the empty container to the depot. Parenthetical numbers in this stage are references to edges in \Cref{fig:6delivery}. The consignee sends its customs clearance and delivery order to the terminal (6.1) and a transfer order to the railway terminal that will take the container (6.2). The railway terminal sends a loading/unloading list to the port terminal (6.3), where internal transportation unloads and loads the appropriate containers. The container and an acceptance document are sent to the railway terminal (6.4a and 6.4c), and a departure notice is sent back to the consignee (6.4b). The consignee sends the required carriage documents to the inland carrier (6.5) which brings the consignee's delivery order to the railway terminal in order to take the container (6.6). The railway terminal gives the carrier a transfer note documenting the interchange (6.7b). The carrier delivers the container to the importer, where it is unloaded (6.7a). The empty container is then brought with the consignee's acceptance order to the depot where it is stored until a new shipment is ready (6.8a and 6.8b).

\Figure[t!]()[scale=1.3, fbox, angle=90, origin=c]{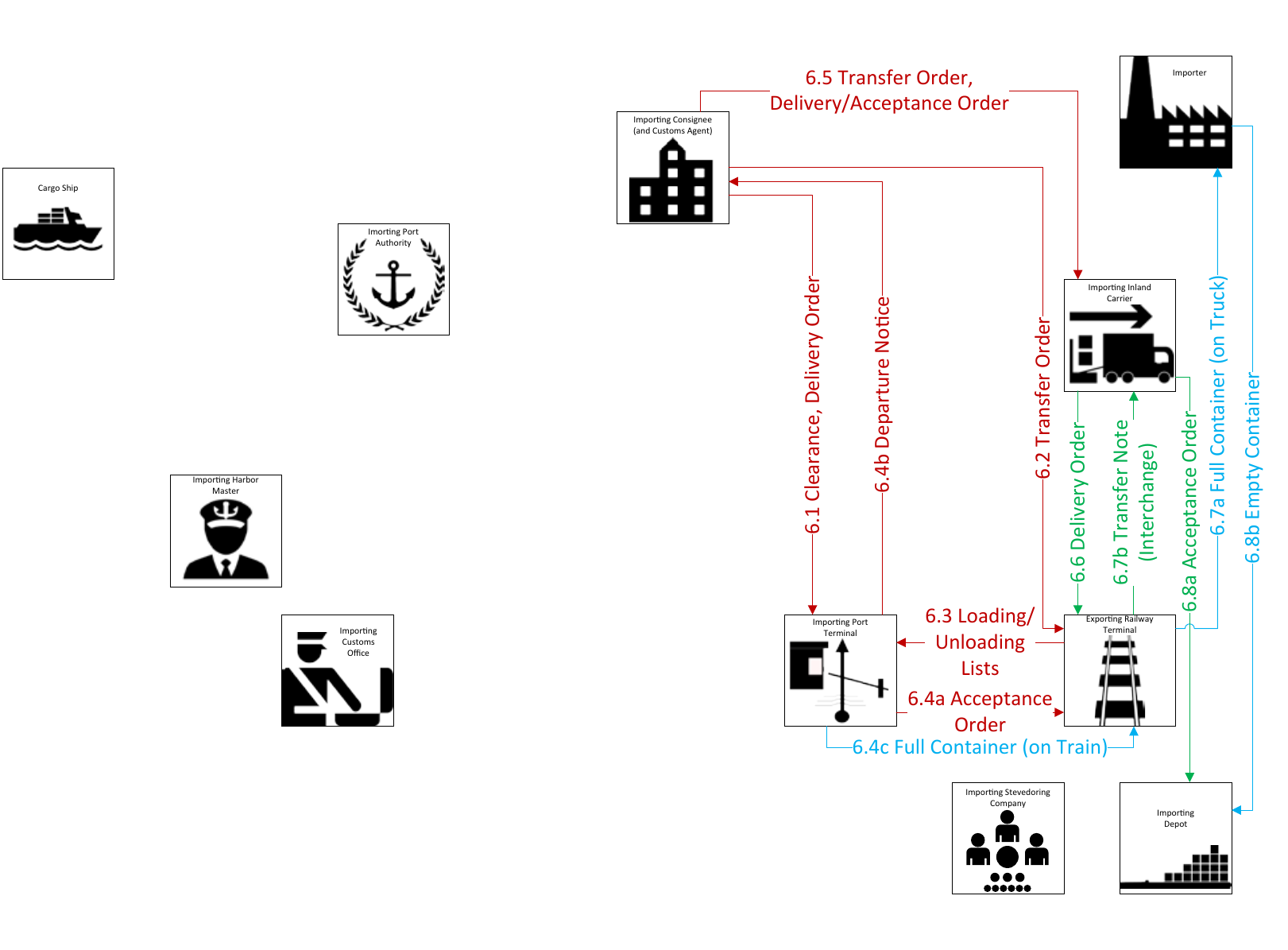}{
    Delivery Logistic Data Flow
    \label{fig:6delivery}
}

\section{In-Depth Vulnerability Assessment}
\label{sec:assessment}

In the previous section, we showed a transactional view of shipping logistics. In this section, we describe the methodology for performing an in-depth vulnerability assessment of two of the modules of the software that controls the transactions previously described. This assessment includes a deep analysis of the software including a low-level code review that goes beyond the use of automated assessment tools. The ultimate goal is to find critical vulnerabilities so that the software providers could remediate them before attackers are able to exploit them.

The modules:
\begin{enumerate}
    \item A web system that facilitates port status and management access for external stakeholders. It also includes services for processing and storing information including ship schedules and location, container locations, gate access status, dangerous goods locations, and loading/discharge lists. External stakeholders, including shippers and consignees, can check the status of this information through this module. This module is 315,000 lines of code, mostly Java and ActionScript.
    \item A web application that communicates yard tractor jobs to the operators in those vehicles. Tractor operators log into the web application from a mobile device. The clients to this module can view the yard tractor jobs and update the status of them as they arrive and are completed. This module is 7,000 lines of code, mostly Java and JavaScript.
\end{enumerate}

The overall effort took 7 person-months. The vulnerabilities found were reported to the head of the development team, followed by several interactions with the development team as to how to fix the vulnerabilities. The patched code was then re-assessed by our team.

Until recently, there was no structured methodology for \textit{in-depth assessment} of software systems at the code level. Simply trying to examine all the code in a complex system such as these would be an overwhelming task, a task beyond any reasonable cost or staffing. Based on our previous experience with analyzing code for security flaws, we developed the First Principle Vulnerability Assessment (FPVA) methodology \cite{kupsch2010first}. FPVA was developed primarily as an analyst-centric approach to assessment, the aim of which is to focus the analyst's attention on the parts of the software system and its resources that are mostly likely to contain vulnerabilities related to high-value assets. FPVA has been used to evaluate many well-known systems, including Google Chrome \cite{miller2013vulnerability}, HTCondor \cite{htcondor}, and Wireshark \cite{miller2013vulnerability}.

Rather than working from known vulnerabilities, the starting point for FPVA is to identify high value assets in a system: those components (for example, processes or parts of processes that run with high privilege) and resources (for example, configuration files, databases, connections, devices) whose exploitation offer the greatest potential for damage by an intruder. From these components and resources, we work outward to discover execution paths through the code that might exploit them. This approach has two immediate advantages. First, it allows us to find new vulnerabilities, not just exploits based on those that were previously discovered. Second, when a vulnerability is discovered, it is likely to be a serious one whose remediation is of high priority.

FPVA starts with an architectural analysis of the code, identifying the key components in a distributed system. It then goes on to identify the resources associated with each component, the privilege level of each component, the value of each resource, the interaction between components, and the delegation of trust. The results of these steps are documented in clear diagrams that provide a roadmap for the last stage of the analysis, which is the manual code inspection. Additionally, the results of this step can also form the basis for a risk assessment of the system, identifying which parts of the system are most immediately in need of evaluation. After these steps, we then use code inspection techniques on the critical parts of the code. Our analysis strategy targets the high value assets in a system and focuses attention on the parts of the system that are vulnerable to not just unauthorized entry but specifically unauthorized entry \textit{that can be exploited}.

After we know where to focus the search, which means after we understand what are the high value assets, we can apply a variety of tools and techniques to the actual analysis of the code. It is worth noting that automated tools complement the manual inspection of the code but never replace it.

In the FPVA of freight ICT systems, we followed the following steps:

\begin{enumerate}
    \item \textbf{Architectural Analysis:} Identify the different software components (processes and threads) running on the different hosts, the communication amongst those components, and the points where the different users interact with the system. Both TOS and PSC are complex, with many components facilitating the interaction among the seaport stakeholders including the port authority, the container terminal, the consignee, and the forwarder.
    \item \textbf{Resource Identification:} Identify the different resources (logical and physical) accessed by the components in step 1. For example, relevant resources include the bill of lading, bayplan, the list of containers with dangerous goods, and the database containing information on the containers on the yard. An attacker gaining access to these critical resources would result in severe damage.
    \item \textbf{Privilege Analysis and Trust Delegation:} Identify the resource protections, the privilege levels at which each component runs, and the delegation of trust. Authentication and authorization of access to resources are also identified in this step. We analyze the trust relationships between key entities such as terminal stations, port operators, forwarders, and shipping companies.
    \item \textbf{Component Evaluation:} Perform a fine-grain evaluation of the critical components and resources identified in step one and two. This step is the most time consuming and involves the identification of vulnerabilities as well as the construction of proof-of-concept exploits. The process of this step is described below.
\end{enumerate}

It is important to emphasize that FPVA helps us to identify vulnerabilities that are not commonly known or described, in addition to common traditional weaknesses. As we mentioned above, steps 1--3 of FPVA identify those parts of the software that would have the highest security impact if they were to be successfully exploited, the \emph{high value assets}. This identification allows us to focus our analyst resources on the parts of the system that are most critical. \hl{For example, consider that after applying steps 1-3, we see that there is a database used to store log entries that are not consumed by any process. So even if an attacker could modify that database, the impact of their action would be null. Now consider a root-owned file whose content is used by a root user ID process. In that case, an attacker gaining access to that file might cause serious harm. The information obtained by the step 3 of FPVA is essential to determine which are the high value resources of the system.} Through this approach, we identified both common vulnerabilities and vulnerabilities specific to the system we analyzed. Examples of vulnerabilities that we found when assessing the TOS are described in the next section, but before that it is worth mentioning examples of common code weaknesses we look when performing a vulnerability analysis \cite{owasp}:

\begin{itemize}
    \item \textit{Improper or insufficient data validation:} refers to accepting and trusting the input supplied by a user without performing validity checks, and is the cause of many types of serious vulnerabilities.
    \item \textit{Improper error handling:} can allow many types of vulnerabilities, including privilege escalation, disclosing information, or denial of service.
    \item \textit{Buffer overflows:} allows a program to overflow the boundary of a memory buffer, either for reading or writing of the member. As a consequence, an attacker can change the behavior of the program or expose sensitive information.
    \item \textit{Numeric errors:} where an arithmetic operation results in a numeric value that is outside of the range that can be represented with a given number of bits, causing the program to make inappropriate decisions that can affect access or modification of the system and data.
    \item \textit{Injection attacks:} these include command injection, SQL injection, and XML injection. Injection attacks occur where a program constructs a string that contains user input (such as their name or address), and then this string is interpreted by the system (such as making a database request). If the program does not limit the use of the user data, it can allow an attacker inappropriate control of the system.
    \item \textit{Web attacks:} cross-site scripting (XSS), cross-site request forgery (CSRF), session hijacking, and open redirect. These attacks can allow an attacker to control or forge access to a website.
    \item \textit{Directory traversal:} a defect where an attacker accesses files and directories that are stored outside their authorized directory in the file system. Such access can expose private information or allow inappropriate access to a system.
\end{itemize}

In this research, we applied the FPVA methodology for the first time in the maritime domain with the goal of making its software less vulnerable to cyber-attack. We applied FPVA to modules of the TOS and PCS provided by a well-known software provider in maritime freight shipping. The next section summarizes our findings.

\section{FPVA Vulnerability Assessment Results}
\label{sec:results}

In this section, we summarize the results of performing an in-depth vulnerability assessment on some modules of a TOS and PCS from a well-known software provider in the domain of maritime freight shipping. A thorough report and discussion of vulnerability results is not within the scope of this paper. It is worth noting that our results were reported to the software developers in full, including close collaboration to remedy the discovered vulnerabilities, and that our team re-assessed the patched software.

We first show an example (\Cref{fig:diagram}) of the artifacts resulting from the first three steps, namely architectural, resource, and privilege analyses, for one of the modules that we assessed. We then briefly describe the vulnerabilities found.

\Figure[bt]()[scale=0.60]{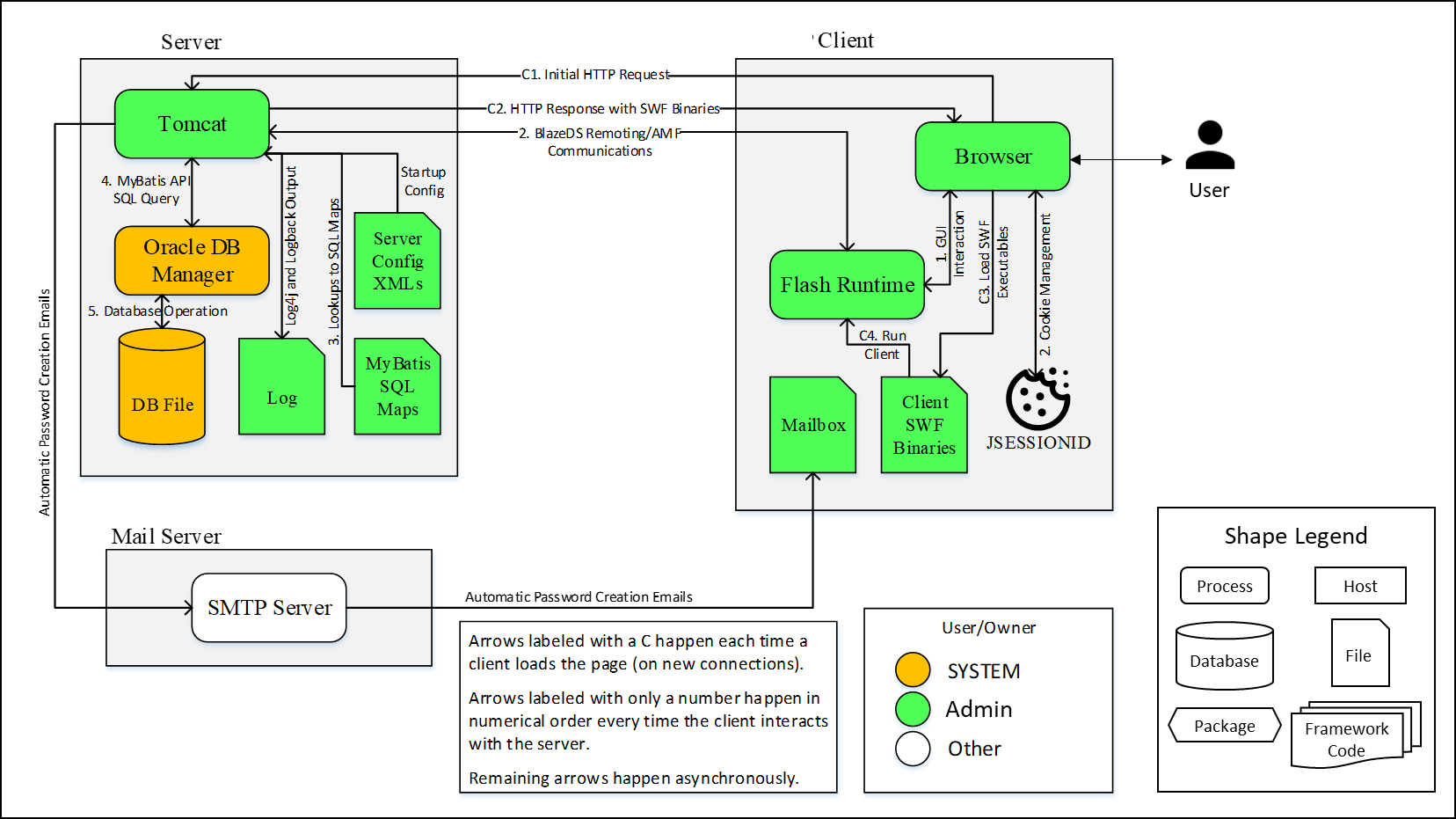}{
    High-level architectural diagram for one of the modules assessed.
    \label{fig:diagram}
}

\Cref{fig:diagram} \hl{shows the \textit{attack surface}, that is the points where the user (or an attacker) can supply input to the system.} It also shows that that module is composed of processes running on three different hosts, each with a different functionality. In orange, we show a process running as user ``SYSTEM'' (the system privileged user), and the resources that are accessed by that process. In green, we show the processes running as user ``Admin'' (the application administrator), and the resources accessed by those processes, such as log files. \hl{The points in the code  where a resource or privilege is abused as a consequence of an attack constitutes the \textit{impact surface}.}  During our analysis we produced detailed low-level diagrams.

\Cref{fig:diagram} \hl{is the result of applying FPVA steps 1-3. In step 4, we inspected the code affecting the identified high-value assets, as we explain next. During our process of understanding component interactions (Architectural Analysis),  access to resources (Resource Analysis), and delegation of operations (Trust and Privilege Analysis), we evaluate for paths through the code from the attack surface to the impact surface. The operations on this path can be described as the cyber kill chain \cite{higgens2013lockheed}, where disrupting any one of those could prevent the attack.

Inspection of the diagram indicates that the most valuable resources are accessed through interactions with the monolithic Oracle database file. This guides our source code evaluation to begin at the code which interfaces with the database. We first enumerate each call to the DB manager in the Tomcat source code, prioritizing accesses to high value tables like the one storing passwords. For each of these calls, we set breakpoints in the code, and then interact with the application normally to record stack traces when any of those breakpoints are reached. Each one of these stack traces becomes a possible attack vector, which we investigate in more detail. For example, with a breakpoint set at every access to the username/password database table, we interact with the system's login, registration, password change, and similar actions. The resulting stack traces provide a list of functions to check for faulty logic or other security flaws that we can exploit. One such function was a unique code path for updating a password which bypassed the normal code path to verify passwords. In this way, we uncovered a vulnerability. }

In our code assessment, we found several high-impact vulnerabilities. Some of the vulnerabilities we found and reported include the following weaknesses:

\begin{enumerate}
    \item Improper authorization and authentication design allowed illegal access to the system's database. Therefore, the following issues arose:
        \begin{itemize}
            \item \textbf{Any user could change any other user's password.} By circumventing client-side validation, an attacker could request a password change for another user without providing a correct current password. This vulnerability was a result of faulty validation logic on the server.
            \item \textbf{Users could access unauthorized services by tampering with client-supplied request metadata.} For example, an attacker could craft a request for Service A with metadata that indicated Service B. The server would authorize the request based on the metadata indicating Service B, but then invoke Service A. This is an example of a trust boundary violation; the server is trusting that the metadata from the client is consistent with the service request's destination. Since client applications can easily be replaced or compromised, the server must assume it is untrusted. For this reason, any validation, authorization, or authentication performed by the client must also be rechecked by the server.
        \end{itemize}
        Design issues such as these are often the most expensive and time-consuming to fix. Some design problems could be detected early in the software development life cycle by using Microsoft's Threat Modelling tool \cite{microsoft2018modeling}. Nevertheless, at this point is it worth quoting ``security systems design is making promises which poor software development practises cannot keep'' \cite{beer2018path}. Complex design problems are only detectable by an expert analyst.
    \item Improper validation in custom file services allowed any user to modify or delete files throughout the server's file system. An attacker could generate a legitimate file download request using the client's user interface and then modify it to specify deleting, downloading, or overwriting any specific file on the server. This vulnerability was a result of both improper sanitizing of the file name to prevent path traversal and lenient access control for the i/o services. Note that the combination of this weakness along with the password compromise vulnerability in weakness 6 would allow an attacker to steal the username and password for every user of the system.
    
        This vulnerability was challenging to find, and it is unlikely to have been found either by automated assessment tools or by penetration testing. The code actually tried to sanitize the input, but it did not cover the specific case that we used for the attack. \hl{Note that the weakness that allowed this vulnerability was previously unknown. It required examination of the code to discover it. As a result, this would not have been found with penetration testing. However, now that this is a known technique for attack, new penetration studies can benefit from its discovery.}
    \item A web server did not check client authorization on all requests. Therefore, many operations were vulnerable to unauthorized access, once the user submitted a correct username and password. By not tracking any login state, the server trusts the client to ensure that unauthorized requests are not made. This is a violation of the trust boundary between client and server.
    
        This vulnerability was challenging to find and required a careful inspection of the code. Neither automated assessment tools nor penetration testers are likely to have discovered it.
    \item An attacker could arbitrarily add log entries to log files. By doing that, the attacker could erase log file history in 2 minutes, as when a threshold was met, the oldest log file was deleted. This vulnerability alone is not severe; however, it may allow an attacker to hide other dangerous activities by overwriting the log.
    
        The effort to find this type of vulnerability is medium: it will not be found by automated assessment tools, but it might be found by penetration testers.
    \item HTTP traffic was not encrypted. As a consequence, the system was vulnerable to:
        \begin{itemize}
            \item \textbf{Session hijacking:} HTTP sessions are tracked using session ID cookies. The server determines client identity and state by associating data with a particular session. If traffic is unencrypted, the value of this session ID can be recorded by an attacker. The attacker can then send requests using that session ID to effectively impersonate the victim, gaining access to all resources available to the victim whose session was hijacked.
            \item \textbf{Password sniffing:} A user's username and password is transmitted in plain text when logging into the system. Any devices connected to the same physical (or virtual) network as a client or server will be able to read the username and password of any user that logs into the system via that network.
            \item \textbf{Sensitive information exposure:} Because all system traffic is unencrypted, an attacker can observe all of the transactions and requests made to the system without directly accessing the system. For example, if a port administrator requested a schedule of dangerous goods while connected to a public network, then any device on that public network could also view that schedule.
        \end{itemize}
        This vulnerability might have been found by automated assessment tools or network monitoring tools.
    \item Password compromise: Instead of using a salted, one-way, cryptographic hash function, the system stores passwords using an insecure form of two-way encryption. The function uses the decryption key as a password's initialization vector, storing this key in both the database and configuration files. The server also writes the encryption key to the general server log every time a password is checked or updated. In the case of a stolen or compromised database file (which was made possible by weakness 2), an attacker could trivially decrypt the passwords stored in the database. This would lead to full compromise of all accounts and disclosure of users' (potentially reused) passwords.
    \item Use of vulnerable versions of third-party software components exposed the system to existing exploits for those components. In any modern software system, third-party components such as framework libraries, operating systems, compilers, and protocols make up a large part of the \textit{software supply chain}. Many of these components contain dangerous vulnerabilities that may compromise the systems depending on them. The presence of dynamic dependencies and non-standard update channels make it difficult to track vulnerable components. 
    \hl{It is definitely recommended to check for CVEs \cite{mitre_cve}. CVEs are Common Vulnerabilities and Exposures, a MITRE-curated catalog of announced vulnerabilities in software systems and components that show existing vulnerabilities in the third-party packages and frameworks on which your software depends. CVEs can also be found in the US NIST National Vulnerability Database (NVD) \cite{nist2020nvd}. Tools such as OWASP Dependency Check can help with this task \cite{owasp_dependency}. Nevertheless, it should be kept in mind that the interaction of secure components can result in a vulnerability, and those are not likely to be detected by automated tools.}

\end{enumerate}

\hl{During the assessment, every time we found a vulnerability, we reported it to the head of the development team. We jointly discussed the mitigations and re-assessed the mitigated version of their software. This process sometimes took several iterations. It is worth mentioning that the owners of the software addressed the security issues extremely promptly. The cost associated with fixing the software was small as compared to the prohibitive cost of being victim of a cyber-attack.}
\section{Conclusions}
\label{sec:conclusion}

For this project, we formed a key collaboration between an experienced academic cybersecurity team and a well-known commercial software provider that manages maritime shipping. We started with a detailed study of the electronic (and paper) information flow involved in maritime freight shipping, highlighting the cyber components involved in this domain. From this study, it was clear that electronic information dominated these processes, and that the ICT systems involved are critical to safe and timely deliveries of shipments.

In addition, we showed a critical gap in the evaluation of the security of these ICT systems. While there have been useful risk assessments of ports, including identifying cybersecurity as a key area of risk, these assessments did not go on to evaluate the software for actual vulnerabilities. For what we believe it is the first time, we conducted a deep dive software vulnerability assessment of some of the critical modules of the TOS and PCS provided by Total Soft Bank. To do that, we applied the First Principles Vulnerability Assessment (FPVA) methodology to those systems and found several significant vulnerabilities in the code. Most of these vulnerabilities would not have been found by the more common practices of using software scanning tools or black-box penetration testing.

Our study provided strong evidence that the shipping domain would benefit from more in-depth software vulnerability assessments, whether it is motivated by regulation, stakeholder trust, or other means.

Total Soft Bank, who allowed their software to be used for this assessment, has taken a significant step forward in providing the maritime shipping industry with a model for more secure ICT infrastructure. This is only a first step, and we hope to see this work extended to other vendors and other aspects of maritime shipping. The goal is to address this problem in a global way.

We believe that this work could provide the foundation for recommendations and guidelines for the maritime freight shipping sector on securing the code of their ICT systems

\bibliographystyle{IEEEtran}
\bibliography{references}

\begin{IEEEbiography}[{\includegraphics[width=1in,height=1.25in,clip,keepaspectratio]{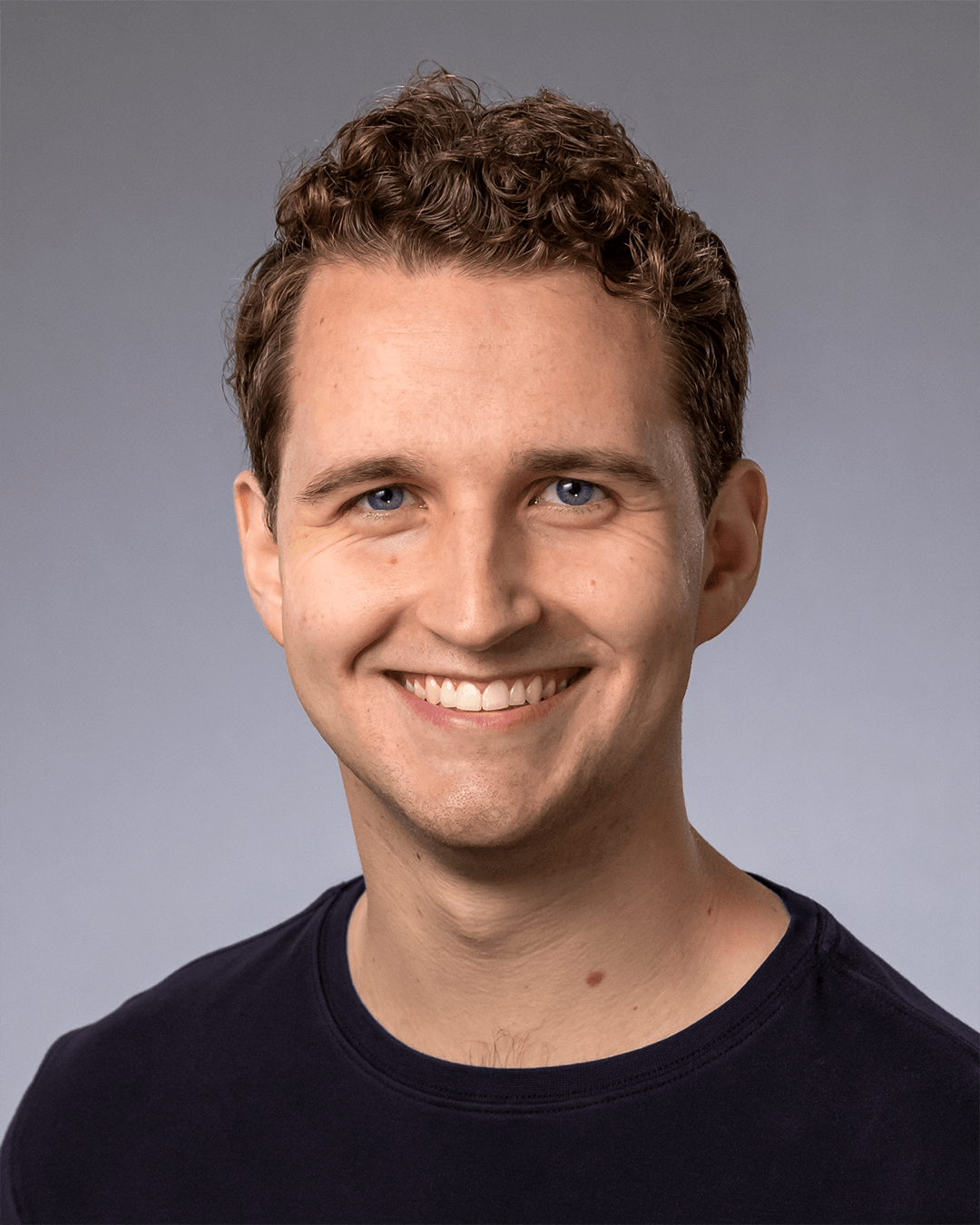}}]{Joseph O. Eichenhofer}
earned his bachelor's degree in computer engineering and computer science from University of Wisconsin-Madison in 2018, where he assisted research efforts related to the software vulnerability assessment project. He also contributed to the creation of a new undergraduate-level course on secure programming. In 2020, he graduated with his M.S.E. in computer science, completing a thesis to design a novel technique for improving censorship resistant communication systems. He also led an effort to redesign parts of the department's undergraduate-level information security course. He is now a production security engineer for Dropbox, Inc. in San Francisco.
\end{IEEEbiography}

\begin{IEEEbiography}[{\includegraphics[width=1in,height=1.25in,clip,keepaspectratio]{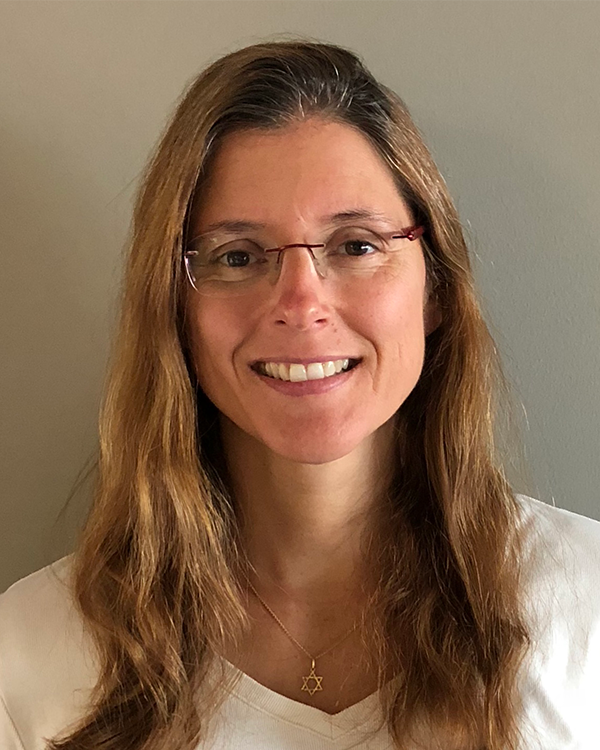}}]{Elisa Heymann}
is a Senior Scientist at the NSF Cybersecurity Center of Excellence at the University of Wisconsin-Madison and an Associate Professor at the Autonomous University of Barcelona. She was also in charge of the Grid/Cloud security group at the UAB and participated in major Grid European Projects. Dr. Heymann's research interests include software security and resource management for Grid and Cloud environments. Her research is supported by the NSF, Spanish government, the European Commission, and NATO.
\end{IEEEbiography}

\begin{IEEEbiography}[{\includegraphics[width=1in,height=1.25in,clip,keepaspectratio]{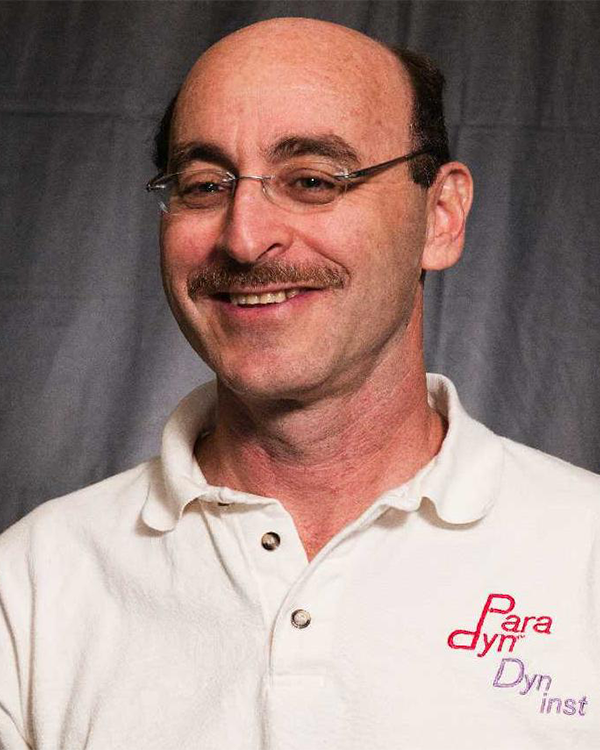}}]{Barton P. Miller}
is the Vilas Distinguished Achievement Professor and Amar \& Belinder Sohi Professor of Computer Sciences at the University of Wisconsin-Madison. He is Chief Scientist for the DHS Software Assurance Marketplace research facility and is Software Assurance Lead on the NSF Cybersecurity Center of Excellence. In addition, he co-directs the MIST software vulnerability assessment project in collaboration with his colleagues at the Autonomous University of Barcelona. He also leads the Paradyn Parallel Performance Tool project, which is investigating performance and instrumentation technologies for parallel and distributed applications and systems. His research interests include systems security, binary and malicious code analysis and instrumentation extreme scale systems, parallel and distributed program measurement and debugging, and mobile computing. Miller's research is supported by the U.S. Department of Homeland Security, U.S. Department of Energy, National Science Foundation, NATO, and various corporations. In 1988, Miller founded the field of Fuzz random software testing, which is the foundation of many security and software engineering disciplines. In 1992, Miller (working with his then-student, Prof. Jeffrey Hollingsworth), founded the field of dynamic binary code instrumentation and coined the term ``dynamic instrumentation''. Dynamic instrumentation forms the basis for his current efforts in malware analysis and instrumentation. He is a Fellow of the ACM.
\end{IEEEbiography}

\begin{IEEEbiography}[{\includegraphics[width=1in,height=1.25in,clip,keepaspectratio]{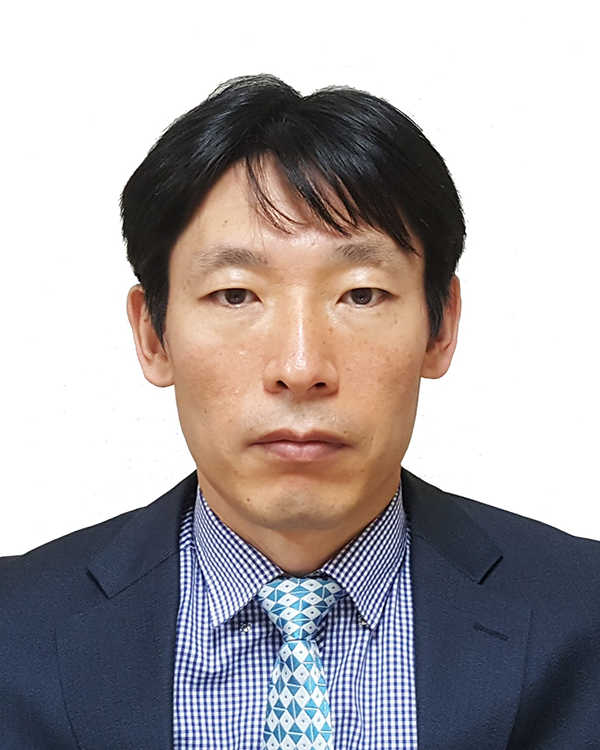}}]{Kyung Won (Arnold) Kang}
is the Director of Total Soft Bank Ltd. He received his B.A. in electronics and communications in 1994 and M.S. in distributed network computing in 2004 from the Korea Maritime University. He has 25 years of experience in the maritime, port, and logistic industry and has led the development of the Terminal Operation System (TOS) and Port Community System (PCS) of Total Soft Bank Ltd.

\end{IEEEbiography}

\EOD

\end{document}